# Effect of wearing a new prophylactic orthosis on postural balance

## Authors

Julien Romain

BV SPORT France

University of Reims Champagne-Ardenne

Mail: j.romain@bvsport.com

Prof. Ahlem Arfaoui

University of Reims Champagne-Ardenne

Laboratory Performance, Health, Metrology and Society (EA 7507)

Mail: ahlem.arfaoui@univ-reims.fr

Prof. William Bertucci

University of Reims Champagne-Ardenne

Laboratory Performance, Health, Metrology and Society (EA 7507)

Mail: william.bertucci@univ-reims.fr

## Abstract

Purpose

The purpose of this study is to evaluate the effect of an innovative prophylactic knee orthosis on postural balance. This prophylactic knee orthosis is designed with a compression that is oriented in a chosen direction. The purpose of this compression is to improve stability in dynamic situations. Orthoses are used to provide functional improvements to knee problems. However, more scientific validation is needed for this type of product.

Methods

20 sportsmen in team sports performed a functional test: the Y-Balance Test. This reliable and reproducible test allows to evaluate the postural balance of the lower limb. The subjects were tested in 3 conditions: prophylactic orthosis with innovative compression, control orthosis (with no compression) and without orthosis. The average of the three trials were collected in each direction and condition.

Results

The prophylactic orthosis had a better standardized score in the anterior direction ($p<0.05$) and a better composite score ($p<0.05$) than the control orthosis (no compression). However, there were no differences in the normalized score in the other directions. There were no significant differences between the prophylactic orthosis and without orthosis.

Conclusion

Wearing the prophylactic orthosis improves postural balance compared to a orthosis with no compression. But there is no difference between the prophylactic orthosis and without orthosis on postural balance.

**Keywords**

## I.  Introduction

The sports market is flooded with products that seek to help athletes improve their performance or reduce the risk of injury. Knee braces or knee orthoses were initially intended to immobilize a joint outside of physical activities. Thereafter, they are democratized to support, maintain or assist the knee that has lost some or all of its functional capabilities. They also allow to reduce pain, to increase physiological performance and to have a proprioceptive effect (Baron, 2016).

The influence of compression sleeves (calf) was evaluated on proprioception-related accuracy in a knee repositioning task with and without compression (Ghai et al, 2018). The results showed that calf compression sleeves can improve knee proprioception. Various studies have also sought to evaluate the importance of knee brace compression on agility and neuromuscular control (Bodendorfer et al, 2019) or also postural control (Baige et al, 2020). Like the calf sleeves the knee pads must respect a very strict dosage of pressure in order not to hinder the practice or create a tourniquet effect.

One important parameter that has not yet been discussed is the impact of knee braces on stability in specific tasks. Knee braces also have a significant effect on knee control during dynamic tasks (step down, single leg drop jump and jump with a half turn). Results validated on subjects with previous anterior cruciate ligament (ACL) injuries (Hanzlikova et al, 2019). The knee brace significantly influences knee kinematics. In contrast the reliability of knee braces is to be qualified according to the type of exercise performed and the nature of the injury obtained. A prophylactic orthosis, without rigid reinforcements, does not have the same impact on the stability of movement as a rehabilitation orthosis, with rigid reinforcements. This case could be observed with a study for lunge movements (Bodendorfer et al, 2019). The immediate effect of the knee brace is limited to the control of tibial rotation for post ACL injury athletes during a lunge exercise. It has also been shown that knee orthoses can provide functional improvements related to the knee joint (Sharif et al, 2017). However, additional work is needed to validate this hypothesis due to the lack of consistency and rigor of the studies analyzed.

The hypotheses of this study are that the knee brace, designed with an innovative compression base, should improve dynamic balance during a Y Balance Test. Its composite score should be higher than without the brace. It would also be interesting to observe the possible differences according to the different directions, in relation to the characteristics of the orthosis. This orthosis should limit postural instability, especially during movements that change direction.

## II. Materials & Methods

### II.1. Population

The study was conducted in 20 students sportsmen (in team sport with at least 3 training sessions per week) including 7 women and 13 men (Table 1) in this twenties. A history of clinical ankle sprain was an exclusion criterion for participation in the study. Each subject completed a consent and approval form to participate in this study.

*Table 1: Anthropological measures of the test subjects*

| | Age | Mass (kg) | Size (cm) | |
|---|---|---|---|---|
| | $21.9 \pm 1.4$ | $66.8 \pm 8.3$ | $167.4 \pm 20.7$ | |
| **Thigh length (cm)** | **Leg length (cm)** | **Calf circumference (cm)** | **Thigh circumference (cm)** | **Knee circumference (cm)** |
| $49.4 \pm 2.3$ | $48.4 \pm 2.8$ | $36.9 \pm 2.7$ | $52.5 \pm 4.9$ | $36.3 \pm 2.4$ |

### II.2. Materials

A functional test has been chosen to highlight the postural balance of the lower limb: the Y-Balance Test (FMS; Functional Movement Systems Inc, Chatham, VA; Fig. 1). It is a derivative of the Star Excursion Balance Test with only 3 directions, or branches, in relation to the position of the supporting foot: anterior (ANT), posterolateral (PL) and posteromedial (PM) (Pliksy et al, 2009). This test has been shown to be reliable and reproducible for assessing postural balance of the lower limb (Baron, 2016; Picot et al, 2012; Plisky et al, 2009). In contrast it is essential to master the protocol and its standardization in order to obtain reliable and reproducible results. Indeed, a bad positioning of the foot, a bad placement of the hands or a too important number of passages can lead to measurement errors that can affect the results. According to several studies, it is necessary to perform 4 training trials in order to limit the learning factor during the test (Baron, 2016; Gribble et al, 2012; Olmsted et al, 2002).

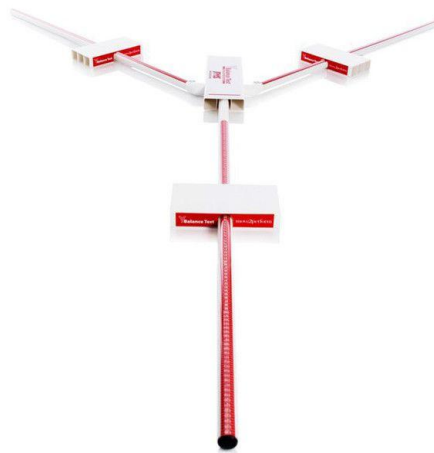

*Fig. 1: Y-Balance Test*

This study will compare the postural balance between 3 conditions of orthosis wearing: without orthosis (WITHOUT), with compression orthosis (COMP) and with a control orthosis (CONT). COMP have different level of compression on specific zones around the knee articulation. CONT has the same size as COMP but no compression applied. CONT is the type of orthosis available that have no compression around the knee. CONT and COMP are visually the same for the experimentation in full black but the texture of COMP is different because of the compression zones.

II.3. <u>Protocole</u>

The tests will be conducted in crossover, controlled and randomized. Subjects will be asked to place the foot in the center of the apparatus according to the recommendations of Picot et al (2014), in the form of a Y (fig. 1) in monopodal support. The hands will always be placed at waist level. They will then have to extend one leg as far as possible in the 3 directions previously mentioned: ANT, PL and PM. Each time the foot is placed, the distance is measured between the position of the foot and the wedge (fig. 1). The 3 trials were then averaged and reported to the length of the limb (combination of tight length and leg length; Table 1) and per direction to obtain a standardized score (%):

$$Score\ normalised\ X\ (\%) = \frac{Average\ of\ the\ 3\ trials\ (cm)}{Length\ of\ the\ lower\ limb\ (cm)} \times 100$$

A composite score (%) was determined using each direction and sense:

$$Composite\ score\ (\%) = \frac{ANT\ (\%) + PM(\%) + PL(\%)}{3}$$

After 4 training trials (Gribble et al, 2012; Omlsted et al, 2002), the test was performed 3 times in a randomized manner in each orthotic condition (WITHOUT, CONT, and COMP) and in clockwise or counterclockwise direction only on the strong leg. The strong leg is either known by the subject or it is determined by a simple test. This test consists of standing behind the subject, who must stand straight and with his feet together. The subject is pushed in the back. The subject loses his balance and uses his strong foot to recover (Schorderet et al, 2021).

II.4. <u>Statistical analysis</u>

The study data were analyzed using normality (Shapiro and Wilks) and homogeneity of variances (Levene) tests. Then nonparametric descriptive statistics (Friedman) were performed according to normality. The confidence index was set at 95% (Statistica 12, Statsoft).

## III. Results

### III.1. Score normalized to limb length by direction

For the score normalized to limb length by direction (Table 2), there is a significant improvement in the normalized score in the ANT direction (Fig. 2) for the COMP condition compared to the CONT condition (p=0.03). The scores of the COMP condition are slightly higher than the other conditions in the posterior directions, PL (Fig. 2) and PM (Fig. 2), but no significant difference can be noted.

*Table 2: Score normalized to limb length in all 3 directions and Composite Score on the Y-Balance Test*

|  |  |  |  |  | **Percent difference between each condition** | | |
|---|---|---|---|---|---|---|---|
|  |  | COMP | CONT | WITHOUT | COMP x CONT | COMP x WITHOUT | CONT x WITHOUT |
| *Score normalized (%)* | ANT | **61.76 ± 7.35*** | **60.09 ± 6.90** | 61.33 ± 6.52 | +2.8% | +0.7% | +2.1% |
|  | PL | 98.90 ± 10.63 | 97.50 ± 11.17 | 98.87 ± 11.80 | +1,4% | +0.1% | +1.4% |
|  | PM | 99.50 ± 9.49 | 98.54 ± 9.31 | 98.02 ± 10.42 | +1% | +1.5% | +0.5% |
| *Composite score (%)* |  | **86.72 ± 8.49*** | **85.38 ± 8.36** | 86.08 ± 8.83 | +1% | +1.6% | +0,7% |

*p<0.05 between COMP and CONT

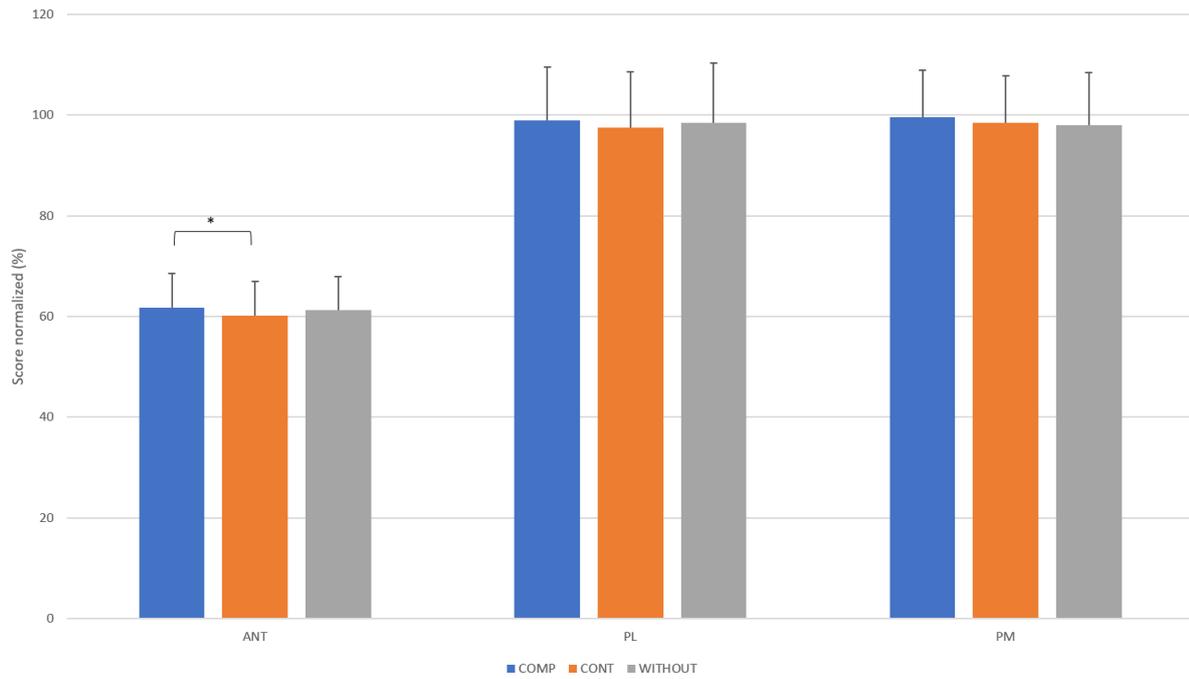

*Fig. 2: Score normalized to limb length in all 3 directions (blue: COMP; orange: CONT;*

*green: WITHOUT; \*p<0,05)*

### III.2. Composite score of the Y Balance Test

For the composite score (Table 2), there was a significant improvement in the composite score of the COMP condition (Fig. 3) compared to the CT condition (p=0.02). The composite score of the COMP condition is also higher than that of the WITHOUT condition, but there is no significant difference to note.

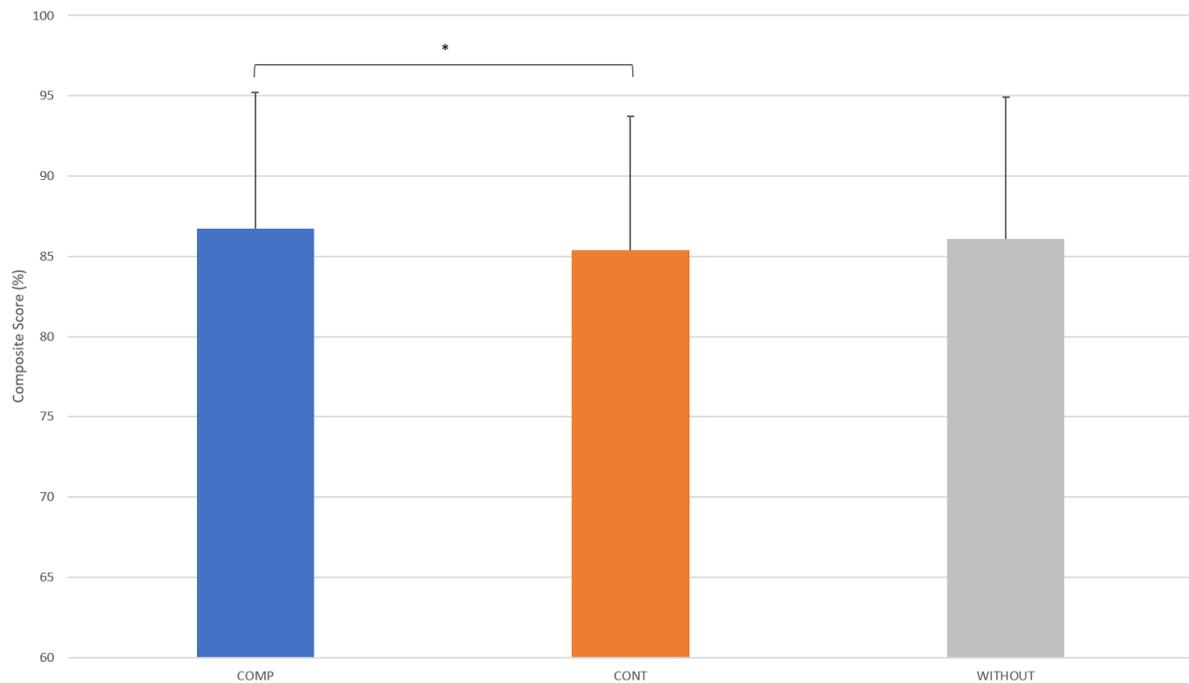

*Fig. 3: Composite Score on the Y-Balance Test (blue: COMP; orange: CONT; green: WITHOUT, \*p<0,05)*

## IV. Discussion

The purpose of this study was to evaluate the effect of a prophylactic knee brace on postural balance. Postural balance was evaluated with the Y-Balance Test, a functional test of the lower limb. The tests carried out showed that the COMP condition did not significantly improve postural balance compared to WITHOUT condition. Similar results were observed in a study (Choi et al, 2020) that sought to evaluate the contribution of compression and taping (silicone elastic band) on postural balance, notably with a Y-Balance Test. They showed that there were no significant differences between wearing a compression product and no compression on the knee for postural balance. This observation may be due to several parameters such as the technique of the knee brace (Fig. 5) as well as the choice of the subjects. Is it possible that the placement of the compression around the knee are not good enough to have an impact on the stability. Or maybe the technique of compression is not adapted for this use. Then the subjects are healthy athletes who do not use knee orthoses.

These healthy subjects do not have postural balance or instability problems that may be related to knee injuries. They do not feel the need to apply control to their joint. In the previous studies cited, the subjects all had a history of knee joint damage such as cruciate ligament rupture. More studies are necessary to develop these assumptions.

The knee braces in the sports market are essentially made with metal reinforcements but they are not allowed in all sports competitions. The knee brace with metal reinforcements has the possibility to restrict or to block articulation. Unlike the prophylactic brace (Fig. 5a) in this study, which does not have metal reinforcements, it has an innovative compression that does not exist on the sports market. The brace studied in this study has specific and innovative compression zones. Its structure of mesh and pressure allows an orientation of the pressures around the knee joint. This compression on COMP allows the pressure to be directed in one direction in order to reproduce a strapping or kinesio taping type of support. This action aims to improve dynamic balance. Kinesio taping placed around the knee improves dynamic balance during a Y-Balance Test or on knee positioning error (Saki et al, 2022).

The importance of population characterization for postural control was studied (Baige et al, 2020). The improvement of postural control with a calf compression sleeve was evaluated, especially the inter-individual variability. Analysis of the results by group did not show significant results. However, the inter-individual analysis of the results showed that subjects with good stability without compression had no effect when wearing the compression. On the other hand, subjects who did not have good stability without compression showed a significant improvement in postural control with compression. It is therefore interesting to make pre-selections of subjects to determine a group of people with postural instability. A study had similar results on the impact of a compression knee brace or taping on balance and muscle activation before and after fatigue (Cavanaugh et al, 2016). The results showed no significant effect of compression or taping on immediate (pre-fatigue) and post-fatigue balance.

The other interesting result of the present study was the significant improvement in the composite score for the COMP condition compared to the CONT condition ($p<0.05$). The initial hypothesis assumed that the COMP condition would result in a significant improvement in dynamic balance compared to the WITHOUT condition and thus the CONT condition. This result first shows that there is no placebo effect. The visuals of the COMP and CONT orthoses (Fig. 5) are very similar despite the mesh effects on COMP. This case could be observed in several studies (Higgins et al. 2009; Cavanaugh et al. 2016). Therefore, it can be inferred that it is important to choose a prophylactic brace with compression than a simple brace without compressive effect (such as a tube). In addition, the learning phenomenon was limited with the 4 trials and with the randomization of the running order of each condition. This also eliminates learning bias (Higgins et al. 2009; Gribble et al. 2012).

The nuance that can be brought to the results in the literature comes from the type of brace, knee brace or compression applied, as there are different types of compression: degressive, progressive, selective or constant. Each of the products has different characteristics and benefits, whether in terms of pressure exerted, reinforcements applied (relaxation of the patella) or materials used (silicone). It is therefore necessary to be cautious when comparing the results obtained in the literature.

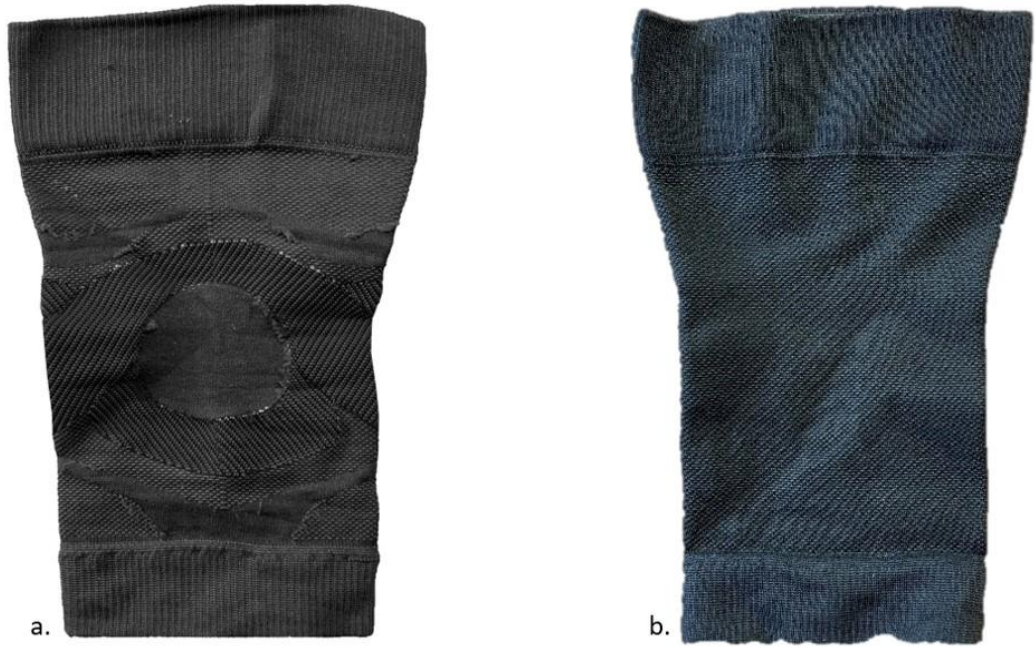

*Fig. 5: Illustration of prophylactic orthosis (a) and control orthosis (b)*

In future studies, it would be interesting to study the impact of this prophylactic orthosis, particularly during changes of direction or receptions. These movements are elements that athletes encounter and can increase the instability factor. This factor is even more important for athletes who have already had an ACL injury or reconstruction. The impact of orthotics on stability during this specific type of task has been discussed (Hanzlikova et al, 2019), but the orthotics studied had different characteristics than in this study. They had shown that the orthosis significantly influences the kinematics of the knee.

## V. Conclusion

The purpose of this study was therefore to evaluate the effect of a new prophylactic knee brace on postural balance. The tests performed with the Y-Balance test showed that the prophylactic orthosis (COMP) improves dynamic balance compared to the CONT condition. Therefore, a knee brace should have targeted and oriented compression zones to improve postural balance. It is necessary for people opting for an orthosis not to choose a simple orthosis without compression (a tube) to feel an improvement in dynamic balance.

The other point to remember is the importance of using a prophylactic brace for healthy athletes with no balance problems or history of knee injuries. Selection of subjects with a history of knee injuries would be a more appropriate choice for this type of experiment.

To further investigate these results, the next step is to perform these tests with a group of subjects with postural instability in order to verify the impact of the prophylactic orthosis on this specific population.

<u>Conflicts of interest</u>

The author, Julien Romain, is doing a CIFRE (Industrial agreement for training through research) thesis with the laboratory Performance, Health, Metrology and Society (EA 7507, University of Reims Champagne-Ardenne, France) in collaboration with the company BV SPORT (Saint-Etienne, France). The Prophylactic Knee Orthosis (COMP) was realized by Julien Romain with BV SPORT. The co-authors, Mrs Ahlem Arfaoui and William Bertucci (Laboratory PSMS, University of Reims Champagne Ardenne, France), have no link with the company BV SPORT.

**References**

Baige K, Noé F, Bru N, Paillard T. « Effects of Compression Garments on Balance Control in Young Healthy Active Subjects: A Hierarchical Cluster Analysis ». *Frontiers in Human Neuroscience* 14 (November 2020): 582514.

Baron D. « Intérêt des orthèses dans la prise en charge de la gonarthrose ». *Revue du Rhumatisme Monographies* 83, n⁰ 2 (April 2016): 119-26.

Bodendorfer BM, Arnold NR, Shu HT, Leary EV, Cook JL, Gray AD, Guess TM, et Sherman SL. « Do Neoprene Sleeves and Prophylactic Knee Braces Affect Neuromuscular Control and Cutting Agility? » *Physical Therapy in Sport* 39 (September 2019): 23-31.

Cavanaugh MT, Quigley PJ, Hodgson DD, Reid JC, et Behm DG. « Kinesiology Tape or Compression Sleeve Applied to the Thigh Does Not Improve Balance or Muscle Activation Before or Following Fatigue ». *Journal of Strength and Conditioning Research* 30, n[o] 7 (July 2016): 1992-2000.

Choi NH, et Hwang S. « Relationship between Kinesiotaping and Compression Wear for Postural Balance in Healthy Men: A Cross-Sectional. *Physical Therapy Rehabilitation Science* (December 2020); 9:275-80.

Chuang SH, Huang MH, Chen TW, Weng MC, Liu CW, et Chen CH. « Effect of Knee Sleeve on Static and Dynamic Balance in Patients with Knee Osteoarthritis ». *The Kaohsiung Journal of Medical Sciences* 23, n[o] 8 (August 2007): 405-11

Ghai S, Driller MW, Masters RSW. « The Influence of Below-Knee Compression Garments on Knee-Joint Proprioception ». *Gait & Posture* 60 (February 2018): 258-61.

Gribble, PA, Hertel J, Plisky P. « Using the Star Excursion Balance Test to Assess Dynamic Postural-Control Deficits and Outcomes in Lower Extremity Injury: A Literature and Systematic Review ». *Journal of Athletic Training* 47, n[o] 3 (May 2012): 339-57.

Hanzlíková, I, Richards J, Hébert-Losier K, Smékal D. « The Effect of Proprioceptive Knee Bracing on Knee Stability after Anterior Cruciate Ligament Reconstruction ». *Gait & Posture* 67 (January 2019): 242-47.

Hanzlíková, I, Richards J, Tomsa M, Chohan A, May K, Smékal D, Selfe J. « The Effect of Proprioceptive Knee Bracing on Knee Stability during Three Different Sport Related Movement Tasks in Healthy Subjects and the Implications to the Management of Anterior Cruciate Ligament (ACL) Injuries ». *Gait & Posture* 48 (July 2016): 165-70.

Higgins T, Naughton GA, Burgess D. « Effects of Wearing Compression Garments on Physiological and Performance Measures in a Simulated Game-Specific Circuit for Netball ». *Journal of Science and Medicine in Sport* 12, n[o] 1 (January 2009): 223-26.

Jalali M, Farahmand F, Esfandiarpour F, Ali Golestanha S, Akbar M, Eskandari A, Mousavi SE. « The Effect of Functional Bracing on the Arthrokinematics of Anterior Cruciate Ligament Injured Knees during Lunge Exercise ». *Gait & Posture* 63 (June 2018): 52-57.

Mohd S, Aishah N, Goh SL, Usman J, Kamarul Zaman Wan Safwani W. « Biomechanical and Functional Efficacy of Knee Sleeves: A Literature Review ». *Physical Therapy in Sport* 28 (November 2017): 44-52.

Olmsted LC, Carcia CR, Hertel J, Shultz SJ. « Efficacy of the Star Excursion Balance Tests in Detecting Reach Deficits in Subjects With Chronic Ankle Instability ». *Journal of Athletic Training* 37, n$^o$ 4 (December 2002): 501-6.

Picot B et Forestier N. « Intérêt du renforcement spécifique des muscles fibulaires sur l'orthèse Myolux ® dans les changements de direction rapide : » *Kinesitherapie in Sport Information*, n$^o$ 1 (2012): 14.

Plisky PJ, Gorman PP, Butler RJ, Kiesel KB, Underwood FB, Elkins B. « The Reliability of an Instrumented Device for Measuring Components of the Star Excursion Balance Test ». *North American Journal of Sports Physical Therapy: NAJSPT* 4, n$^o$ 2 (May 2009): 92-99.

Saki F, Romiani H, Ziya M, Gheidi N. « The Effects of Gluteus Medius and Tibialis Anterior Kinesio Taping on Postural Control, Knee Kinematics, and Knee Proprioception in Female Athletes with Dynamic Knee Valgus ». *Physical Therapy in Sport* 53 (January 2022): 84-90.

Schorderet, Chloé, Roger Hilfiker, et Lara Allet. « The Role of the Dominant Leg While Assessing Balance Performance. A Systematic Review and Meta-Analysis ». *Gait & Posture* 84 (February 2021): 66-78.